\documentclass[preprint,12pt]{elsarticle}
\usepackage{amsmath, mathrsfs, latexsym, amssymb, color, graphicx, comment, hyperref, scrhack}
\usepackage{lmodern}
\usepackage{physics}

\newcommand{\dfr}[2]{\frac {\displaystyle #1}{\displaystyle #2}}
\journal{Physica A}

\begin{document}

%\begin{frontmatter}

\title{ Bose system critical dynamics near quantum phase transition}
\author[label,label1]{M.\,G.\,Vasin}
\address[label]{Vereshchagin Institute of High Pressure Physics, Russian Academy of Sciences, 108840 Moscow, Russia}
\address[label1]{Udmurt Federal Research Center UrB RAS, 426067 Izhevsk, Russia}
\author[label2]{V.\,M.\,Vinokur}
\address[label2]{Materials Science Division, Argonne National Laboratory, 9700 S. Cass Ave, Argonne, IL 60439, USA}

\begin{abstract}
	
	We show that the change of the  fluctuation spectrum near the quantum critical point (QCP) may result in the continuous change of critical exponents with temperature due to the increase in the effective dimensionality upon approach to QCP.  The latter reflects the crossover from thermal fluctuations white noise mode to the quantum fluctuations regime. We investigate the critical dynamics  of an exemplary system obeying the Bose--Einstein employing the Keldysh--Schwinger approach and develop the renormalization group technique that enables us to obtain analytical expressions for temperature dependencies  of critical exponents.
\end{abstract}

%\begin{keyword}
%Quantum phase transition, critical exponents, critical dynamics, Bose--Einstein statistics
%\end{keyword}

%\end{frontmatter}

\flushbottom
\maketitle

\thispagestyle{empty}

\section{Introduction}

Quantum phase transitions (QPT) and related quantum criticality is one of the central topics in modern physics, see\,\cite{Girvin1997,Sachdev2011}  for comprehensive reviews. Quantum phase transitions taking place at the zero temperature drven by the variation of one of the characteristic parameters are ubiquitous in nature and occur in the diversity of systems ranging from the rare-earth magnetic insulators\,\cite{5Bitko} and heavy fermion hosting compounds\,\cite{6Coleman, 7Lohneysen} to high-temperature superconductors\,\cite{8Dagotto,9Sachdev} and superconducting films\,\cite{Goldman2010}. Remarkably, near the QPT their dynamic and static criticalities are intricately intertwined at variance to those in classical transitions, and this feature may be the source of still remaining challenges in understanding the details of the crossover from the classical to quantum critical behaviors.

One of these challenges is the possibility of the temperature dependence of critical exponents, which have to change from their classical to their mean-field values upon the decrease in temperature. In spite of tantalizing experimental reports\,\cite{Steijger,Erkelens,Stishov2,Stishov1,Bittar}, its very existence still remains the subject of fierce debate. Likewise, the theoretical description of the possibility of such a change remains an open problem.
Here we try to meet this challenge and develop the approach based on the Keldysh--Schwinger technique enabling the description of classical and quantum critical dynamics within the unique framework accounting for, in particular, for the dissipative processes.

The paper is organized as follows: we shortly touch upon the common wisdom point of view on quantum--classical crossover, and start with introducing the Keldysh--Schwinger approach to the non-equilibrium dynamics description of a Bose system. Then we take the respective classical and quantum limits, and, finally,  we describe the crossover between the dissipative and adiabatic regimes. Having established the technique, we apply it to description of the dissipative quantum regime near the QCP and develop the renormalization group allowing to treat the crossover from classical to quantum fluctuations. We calculate the change of the critical exponents with temperature near the QCP. Finally, the crossover from the dissipative-quantum critical regime to the adiabatic one is explained in terms of renormalization procedure.

\section{Quantum--classical crossover near quantum critical point}

According to common wisdom viewpoint, the quantum critical point is gapless and scale invariant, i.e. the system experiences quantum fluctuations at all frequencies down to zero\,\cite{Girvin1997} fluctuations frequency $\omega_0$. The latter depends on the system nature. For example, in magnet systems this value corresponds to exchange energy. Temperature introduces a new energy scale into the quantum problem and cuts off coherent quantum fluctuations in the infrared spectrum range, $\omega<1/\hbar \beta$, in which thermal fluctuations present.

\begin{figure}[h!]
\centering
   \includegraphics[scale=0.5]{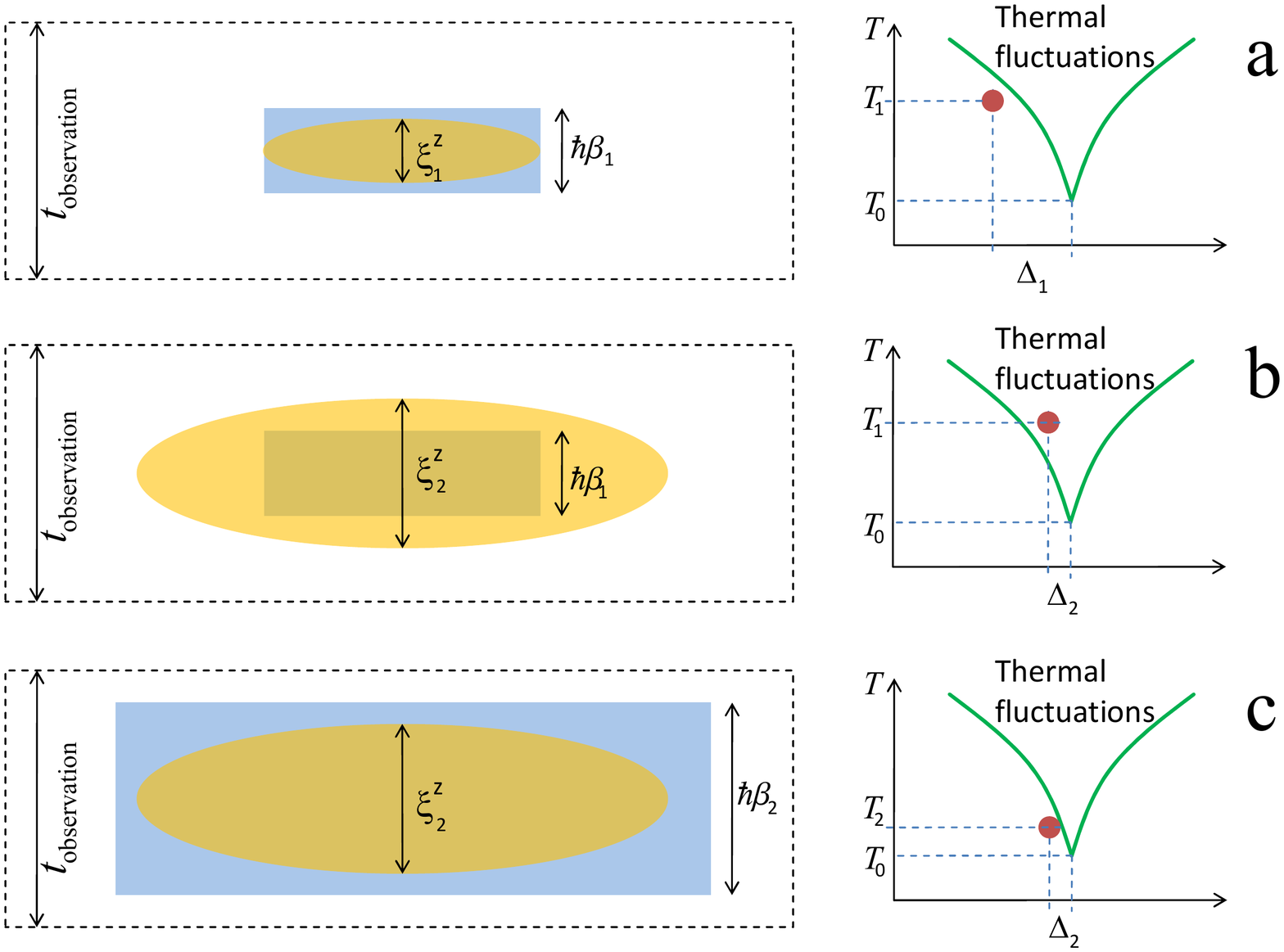}
   \caption{ Illustration of fluctuation regime close to a quantum critical point, inspired by the illustration of the growing correlation volume as the T=0 critical coupling in \cite{Girvin1997}: {\bf a)} the correlation time is shorter than $\hbar\beta_1$, the system is in quantum fluctuation regime; {\bf b)} if the system becomes closely to the critical point, $\Delta_2<\Delta_1$, then the correlation time exceeds $\hbar\beta_1$ and the system goes in thermal fluctuation regime; {\bf c)} the temperature increasing, $T_1\to T_2<T_1$, brings back the quantum fluctuations dominating, $\xi^z=\Delta^{-\nu z}<\hbar\beta_2$. $T_0$ is the minimum limit of temperature corresponding to the zero fluctuation frequency of the system, $T_0=\hbar\omega_0/k_{\mathrm{\scriptscriptstyle B}}$, $k_{\mathrm{\scriptscriptstyle B}}$ is the Boltzmann constant.}
   \label{Fig1}
\end{figure}

We take it that in quantum limit, the effective dimension of $d$-dimensional system is $d+1$, because in addition to the $d$ spatial dimensions one temporal dimension of the size $L_{\tau}=\hbar \beta$ appears. On the other hand, the physics has to be continuous in temperature. As a result, the question arises of how the system ``learns'' that its dimension has been changed. The answer to this is illustrated in Fig.\,\ref{Fig1}, on which one can see how the system can go between different fluctuation regimes. According to this picture, the key parameter depending on the system effective dimension is the ratio of the  correlation time growing near critical point to the quantum temporal size growing with the decreasing temperature. Upon crossing the ratio $\hbar\beta\sim \Delta^{-\nu z}$, the system effective dimension is changes, $d\leftrightarrow d+1$.

As phase transitions are sensitive to system's dimensionality, one can expect that the finiteness of $L_{\tau}$ at $T\neq 0$ will modify the critical behavior as a response to change in $\Delta$ or in temperature. At the longest length scales, the system is now $d$ dimensional. If the correlation time $\xi^z>L_{\tau}$, the critical behaviour corresponds to $d$-dimension system criticality. With temperature decreasing,  $L_{\tau}$ becomes large than correlation time; then the correlated part of the system ``forgets'' about its environment and accordingly about thermal bath. The quantum fluctuations remain only as a result of the effective system dimension becoming $d+1$.

Let us note again that the physics has to be continuous in temperature. Therefore the theory of the quantum phase transition should describe the quantum--classical crossover in dynamics taking into account both quantum and thermal fluctuations, and keep track continuous transition from first ones to the second ones.
In order to satisfy this condition we will build our theory on the dynamic approach to phase transitions.

\section{Keldysh--Schwinger approach to non-equilibrium dynamics description}

The standard framework for the Keldysh--Schwinger technique\,\cite{Keldysh, Schwinger} is the path integrals formulation\,\cite{DF,KO,KAMENEV,Milton}, which for convenience, we briefly introduce here.
Let us consider a quantum many-body system governed by the time-dependent Hamiltonian $\hat H(t)$.
The system's sate at the time moment $t$ is set by the function $|\phi,\,t\rangle$ (see Appendix I).
The time evolution of the system is governed by the evolution operator $\mathcal{\hat U}$: $|\phi,
t+\delta t\rangle=\mathcal{\hat U}_{\delta t}|\phi,\,t\rangle$. This operator evolves according to the Heisenberg equation of motion $\hbar\partial_t\mathcal{\hat U}_{\delta t}=i\left[\mathcal{\hat U}_{\delta t},\,\hat
H\right]$, which is formally solved as
\begin{gather*}
\mathcal{\hat U}_{\delta t}=\exp \left[-\dfr i{\hbar}\delta t\hat H\right].
\end{gather*}

The quantum mechanical amplitude describing the transition from the state $|\phi, t'\rangle $ to the state $\langle \phi, t''|$ is given by (see Appendix II)
\begin{gather}\label{LD}
\langle \phi, t''| \phi, t'\rangle=\langle \phi_{t''}|\mathcal{\hat U}| \phi_{t'}\rangle =
\int \mathfrak{D}\phi\exp \left[i\int\limits_{t'}^{t''}\mathrm{d}t\,\left(\hbar^{-1}\mathcal{L}(\phi)+i\int\limits_V\dfr{\mathrm{d}V}V
\,\phi^*\partial_t\phi\right) \right]
\end{gather}
where $\phi^*$ is the conjugated to $\phi$,
\begin{gather*}
\mathcal{L}(\phi )=\dfr1V\int\limits_{V}\mathrm{d}V\left(\dfr{\mu}2\dot \phi^2-\dfr{\xi}2 (\nabla\phi)^2\right)-P\{\phi\},
\end{gather*}
$\int \mathfrak{D}\phi $ is the functional integration (see Appendix III), $\phi $ is the vector of infinite dimensionality with components $\phi_{\bf r}$, $\mu$, and $\xi$ are the parameters characterising the smoothness of $\phi $ function in space and time respectively, $P\{\phi\}$ is the potential energy, $V$ is the system volume and $\int\mathrm{d}V$ is the integration over this volume.
In the equilibrium case, the initial and final states are in equilibrium with the thermal bath, and are related to the ground state as $ | \phi_{eq}, t \rangle = \sqrt{\rho_0} | 0 \rangle $, where $ \rho_0 (T) $
is the equilibrium density of states. Therefore, $ \langle \phi, \infty | \phi', - \infty \rangle = \langle \phi_{eq}, \infty | \phi_{eq}, - \infty \rangle = \rho_0 \langle 0 | 0 \rangle = \rho_0 $,
i.e. it is a certain constant that depends on the temperature.
This expression can be rewritten as
\begin{gather*}
\langle \phi_{t''}| \mathcal{\hat U}|\phi_{t'}\rangle=
\int \mathfrak{D}\phi \exp \left[i\int\limits_{t'}^{t''}\mathrm{d}t\int\limits_V\dfr{\mathrm{d}V}V\,\phi^* G^{-1}\phi \right],
\end{gather*}
where\,\cite{KAMENEV,PhysRep}
\begin{gather}
G^{-1} = \tau\partial_t^2-c^2\tau \nabla^2 + i\partial_t .
\label{green}
\end{gather}

The quadratic part of (\ref{green}) corresponds to the wave equation $\mu\partial_t^2\phi-\xi \nabla^2\phi =0$, thus $\mu$ and $\xi $ are related by the wave velocity $c$:
$\xi =c^{2}\mu$. The last term in (\ref{green}) describes dissipation.
The time scale in the system described by (\ref{green}), is given by the quantum-mechanical time scale (coherence time), which is defined only the Planck constant and mass $\tau=\hbar^{-1}\mu$. In condensed matter this is the inverse value to the zero-point frequency of the system, $\omega_0=\tau^{-1}$.

We consider a non-equilibrium system interacting with the thermal reservoir and coming to the thermal equilibrium at $t\to \infty$. We assume that at $t=0$, the system is in the out of equilibrium
state $ |\phi_0 \rangle $ and evolves to its final equilibrium state, $\langle \phi_{\infty} |\equiv\langle \phi_{eq} | $, characterized by the equilibrium density of states $\rho _{eq}$. The
transition amplitude is \cite{PhysRep}:
\begin{multline*}
\langle \phi_{eq}, \infty |\phi_0, 0 \rangle = \langle \phi_{eq}|\mathcal{\hat U}_{\infty} |\phi_0 \rangle = \\
\int \mathfrak{D}\phi_{eq}\mathfrak{D}\phi
\exp \left[i\int\limits^{\infty}_{0}\mathrm{d}t\,\left(\hbar^{-1}\mathcal{L}(\phi)+
i\int\limits_V\dfr{\mathrm{d}V}V\phi^*\partial_t\phi\right) +i\hbar^{-1}\tau E_{eq}\int\limits_V\dfr{\mathrm{d}V}V\phi_{eq}^*\phi_{\infty}\right].
\end{multline*}

One can show that this amplitude does not depend on shifting the time $t=\infty$ (see Appendix II). However, it depends on the initial state of the system and, accordingly, on the
choice of the initial time. The averaging operation is not defined in this case and, as a consequence, the statistical theory can not be formulated.

\begin{figure}
   \centering
   \includegraphics[scale=0.47]{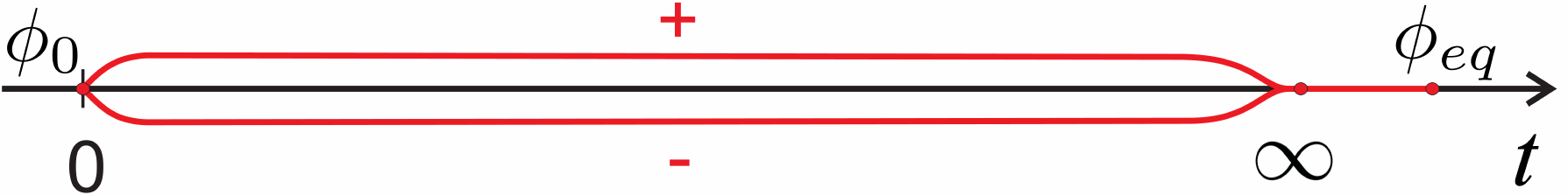}
   \caption{At $t=0$ the present states $\phi_0$ of both fields are equivalent and, therefore, are coherent: $\langle \phi^+_0 |\phi^-_0\rangle =\langle \phi_0 |\phi_0\rangle=1$. At $t=\infty$, the    system reaches the equilibrium state $\phi_{\infty}\equiv\phi_{eq}$ which is also equivalent and coherent for both fields $\phi^+_{\infty}\equiv\phi^-_{\infty}$.  Therefore, the contour is closed for $t=0$ and $t=\infty $. Since at $t>\infty$ the system is in an equilibrium, the two contour branches are equivalent and can be represented by one line.}
   \label{F1}
\end{figure}

To get around this problem, we use the following approach: consider a copy of our system, with the same transition amplitude. We denote the field in the initial system as $\phi^+$ and the field in the
replica system as $\phi^-$. Recall that both fields are, in fact, identical,  hence $\langle \phi_0^-, 0 | \phi_0^+, 0 \rangle = 1 $. Using these two fields, we close the integration contour at $ t = \infty $ (see
Fig.\,\ref{F1}) and write
\begin{gather*}
1\equiv \langle \phi_0^- , 0|\phi_0^+ , 0 \rangle=
\int \mathfrak{D}\phi_{eq}\langle \phi_0^- , 0| \phi_{eq}, \infty \rangle
%\langle \phi_{eq}| \phi_{eq}\rangle
\langle \phi_{eq}, \infty |\phi_0^+ , 0 \rangle =
\mathcal{N}\int \mathfrak{D}\phi^+ \mathfrak{D}\phi^- \times\\
\int \mathfrak{D}\phi_{eq}\exp \left[i\int\limits^{\infty}_{0}\mathrm{d}t\,\left(\hbar^{-1}\mathcal{L}(\phi^+)
-\hbar^{-1}\mathcal{L}(\phi^-)
+i\int\limits_V\dfr{\mathrm{d}V}V\left({\phi^+}^*\partial_t\phi^+
-{\phi^-}^*\partial_t\phi^-\right)\right) +\right. \\
i\hbar^{-1}\tau E_{eq}\int\limits_V\dfr{\mathrm{d}V}V\left({\phi^+}^*(\infty)-{\phi^-}^*(\infty)\right)\phi_{eq}
+i\hbar^{-1}\tau E_{eq}\int\limits_V\dfr{\mathrm{d}V}V\phi_{eq}^2
\Bigg].
\end{gather*}
Here we ``glued'' two branches of the contour at $ t = \infty $, since $ \phi_{eq} $ state does not depend on the choice of the contour. Now the integration over the contour yields unity, and the averaging operation in the system with two fields is well defined because it does not depend on the initial state of the system.

It is convenient to use the frequency representation. Because $E_{eq}\phi_{eq}^2=\tau\int\limits_{-\infty}^{\infty}\hbar \omega \,\rho(\omega )\,\phi ^2_{\omega}\mathrm{d}\omega $, where
$\rho (\omega)=\dfr 12\,\mbox{coth}\left(\hbar\omega/2k_bT\right)$ is the equilibrium density of states, the integration over $\phi_{eq}$ yields
\begin{gather*}
\langle \phi^-_0, 0 | \phi^+_0, 0 \rangle= \mathcal{N}'\int \mathfrak{D}\phi^+\mathfrak{D}\phi^- \exp \left[i\tau^2\int\limits_{-\infty}^{\infty}\mathrm{d}\omega\left\{ \dfr
1{\hbar}\mathcal{L}_{\omega}(\phi^+) -\right.\right.\\  \left.
\dfr 1{\hbar}\mathcal{L}_{\omega}(\phi^-)-\omega\int\limits_V\dfr{\mathrm{d}V}V({\phi^+}^*\phi^+ -{\phi^-}^*\phi^-)-\omega \rho(\omega )\int\limits_V\dfr{\mathrm{d}V}V(\phi^+ - \phi^-)^2\right\} \Bigg].
\end{gather*}

We now perform Keldysh rotation and introduce new fields, $\phi ^{cl}=\sqrt{1/2}(\phi^+ +\phi ^-)$, $\phi ^{q}=\sqrt{1/2}(\phi^+ -\phi ^-)$ ($\phi ^{+}=\sqrt{1/2}(\phi^{cl} +\phi
^{q})$, $\phi ^{-}=\sqrt{1/2}(\phi^{cl} -\phi ^{q})$), which are called ``classical'' and ``quantum'' fields, respectively. Using this rotation, the theory acquires a compact and convenient
form. After Wick rotation $t\to it$ ($\omega\to i\omega$) in the reciprocal space we have
\begin{gather*}
1=\langle \phi^-_0, 0 | \phi^+_0, 0 \rangle= \mathcal{N}'\int \mathfrak{D}\phi^{cl}\mathfrak{D}\phi^{q} \exp \left[-V_{\bf k}^{-1}\iint\limits_{-\infty}^{\infty}\mathrm{d}{\bf k}\mathrm{d}\omega\,{\bar\phi}^*_{k}\hat G^{-1}_{
k}\bar\phi_{-k}\right],
\end{gather*}
where $V_{\bf k}=l^{-d}$ is the system volume in reciprocal space, $\bar \phi=\left\{\phi^{cl},\,\phi^q\right\}$ and $\hat G^{-1}$ is the inverse Green function operator
\begin{equation}
\hat G^{-1}=\tau^3\left[ \begin{array}{cc} 0 & \displaystyle\Box +i \tau^{-1}\omega \\[12pt]
\displaystyle\Box -i \tau^{-1}\omega & \displaystyle \displaystyle\tau^{-1}\omega \,\mbox{coth}\left({\hbar\omega}/{2k_bT}\right) \end{array}\right],
\label{gr}
\end{equation}
where $\Box =(c^2{\bf k}^2 -\omega^2)$ is the  d'Alembert operator describing the evolution of a conservative elastic system.
Accordingly, the Green function operator has the following form:
\begin{equation}
\hat G\equiv \left[\begin{array}{cc}\displaystyle{\hat G}^{\mathrm K} & \displaystyle{\hat G}^{\mathrm R} \\[12pt]
\displaystyle {\hat G}^{\mathrm A} & \displaystyle 0 \end{array}\right]=i\tau^{-2}\left[ \begin{array}{cc}\displaystyle\dfr{\tau^{-1}\omega \,\mbox{coth}\left({\hbar\omega}/{2k_bT}\right)}{\Box^2+ \tau^{-2}\omega^2} & \displaystyle\dfr{1}{\Box -i\tau^{-1}\omega} \\[12pt]
\displaystyle\dfr{1}{\Box +i \tau^{-1}\omega} & \displaystyle 0 \end{array}\right]\,,
\label{gr}
\end{equation}
where ${\hat G}^{\mathrm K}$, ${\hat G}^{\mathrm A}$, and ${\hat G}^{\mathrm R}$ are the Keldysh, advanced, and retarded, Green functions, respectively, and the former satisfies the
quantum fluctuation dissipation theorem (QFDT):
	\begin{gather*}
		[\hat G^{-1}]^{K}=\coth(\hbar\omega/2k_bT)\mbox{Im} \left([\hat G^{-1}]^{A}\right).
	\end{gather*}

\section{Quantum and classical limits of the Keldysh--Schwinger approach}

\subsection{Classical limit}

Let us consider the classical limit so that $T\gg \hbar\omega/2k_{\mathrm{\scriptscriptstyle B}}$. The critical dynamics of the system is determined by the Keldysh element of the Green function matrix, $[\hat G^{-1}]^K=\tau^{-1}\omega \coth(\hbar\omega/2k_{\mathrm{\scriptscriptstyle B}}T)\approx 2k_{\mathrm{\scriptscriptstyle B}}T/\hbar\tau$.  From the non-equilibrium dynamics \cite{V} we know that this element is the correlation function of the external noise acting on the system. Thus, in this case the influence of the thermostat on the system (i.e. the action of the statistical ensemble on its own element) corresponds to the action of the external ``white'' noise, which intensity is proportional to the temperature, and the system is described by the classical non-equilibrium propagator:
	\begin{gather*}
		\hat G^{-1}=\tau^3\left[ \begin{array}{cc}0 & \Box+i\tau^{-1}\omega \\
		\Box-i\tau^{-1}\omega & 2k_{\mathrm{\scriptscriptstyle B}}T/\hbar\tau \end{array}\right].
	\end{gather*}
Then QFDT transforms to the standard Fluctuation Dissipation Theorem (FDT):
	\begin{gather*}
		[\hat G^{-1}]^{K}=({2k_{\mathrm{\scriptscriptstyle B}}T}/{\hbar\omega })\mbox{Im} \left([\hat G^{-1}]^{A}\right).
	\end{gather*}
Here there is only one marginal case case of low energy long wave fluctuations, ${\bf k}\to 0$, $\omega\to 0$,
which is relevant near the critical point. In this limit only the terms with the lowest powers ${\bf k}$
and $\omega $ are to be kept in the Lagrangian.

Hence in the fluctuation region the system is described by the
classical non-equilibrium propagator:
	\begin{gather*}
		\hat G^{-1}=\tau^3\left[ \begin{array}{cc}0 & c^2{\bf k}^2+i\tau^{-1}\omega \\
		c^2{\bf k}^2-i\tau^{-1}\omega & 2k_{\mathrm{\scriptscriptstyle B}}T/\hbar\tau \end{array}\right].
	\end{gather*}
Accordingly, the dispersion relation is $\omega \propto {\bf k}^2$,
therefore the dynamic critical exponent in the first approximation is $z=2$ (for dissipative systems with the non-conserving order parameter).
Now, as usual, going over from statics to the dynamic description implies
using the total (space + time) dimensionality $D=d+z$, where the spatial dimensionality, $d$, controlling the static critical behavior, so that $D=d+2$.  However the ``white'' noise reduces
the effective scaling dimensionality to $D_{eff}=D-2=d$ \cite{Parisi}.
As a result, the critical dimensionalities of the dynamic and static
theories coincide, and the critical behavior of the system is described by
the classical critical dynamics of the  $d$-dimensional system.
We will be referring hereafter to this mode as to the classical critical
dynamics (CCDM), that realizes at $T\gg \hbar\omega/2k_{\mathrm{\scriptscriptstyle B}}$.

\subsection{Quantum case (dissipative dynamics)}

At low temperatures, the system is in quantum fluctuations domain, $\omega\gg 2k_{\mathrm{\scriptscriptstyle B}}T/\hbar$.  If the frequency exceeds or is comparable the inverse time of coherence, $\omega \eqslantgtr\tau^{-1}$, then the inverse Green function operator is presented as
\begin{equation}
\hat G^{-1}=\tau^3\left[ \begin{array}{cc} 0 & \displaystyle\Box +i \tau^{-1}\omega \\[12pt]
\displaystyle\Box -i \tau^{-1}\omega & \displaystyle \displaystyle\tau^{-1}|\omega|  \end{array}\right].
\label{grq}
\end{equation}
The action of the statistic ensemble of the system does not depend on the temperature, and the QFDT assumes the following form:
	\begin{gather*}
		[\hat G^{-1}]^{K}=\mbox{sign}(\omega )\mbox{Im} \left([\hat G^{-1}]^{A}\right),
	\end{gather*}
and the thermostat on the system corresponds to the action of external ``color'' noise with the intensity independent of the temperature and proportional to $\omega$.

This significantly changes the critical properties of the system as compared to those in the classical case.
The total dimensionality now remains $D=d+2$. However, the quantum fluctuations noise in contrast to the thermal ``white'' noise, does not decrease the effective scaling dimensionality \cite{Vasin}, therefore the effective dimensionality of the dissipative quantum system is greater by 2 than its static dimensionality, so that $D_{eff}=D=d+2$. The disagreement between the static and dynamic theories is accounted for by the fact that in the quantum case there is no static limit, and the only correct results are those of the dynamic theory. The corresponding dynamic mode can be referred to as the dissipative quantum critical mode (DQCM).

\subsection{Quantum case (adiabatic dynamics)}

If the quantum limit still holds, but the coherence time is so large that frequency is smaller inverse time of coherence, $2k_{\mathrm{\scriptscriptstyle B}}T/\hbar\ll\omega\ll\tau^{-1}$, the system dynamics is governed by the adiabatic mode in which the dissipation can be neglected. Therefore we let $\tau^{-1} \to 0$, and from Eq.\,(\ref{grq}) find
	\begin{gather}\label{S1}
		\hat G^{-1}\approx \tau^3\left[ \begin{array}{cc}0 & \Box \\
		\Box & 0\end{array}\right],
	\end{gather}
with the dispersion relation becoming $\omega\propto |{\bf k}|$.  Accordingly, the critical dynamic exponent is $z=1$,
and the critical behavior is that of the static system with the effective dimensionality $D_{eff}=d+1$.
Furthermore, the critical behavior of the three-dimensional system is described, in this parameters range, by the mean field Ginzburg--Landau theory, since the effective dimensionality is equal to upper critical dimensionality, $D_{eff}=d_c^+\equiv 4$.
This regime can be referred to as the adiabatic quantum mechanical mode (AQM).

\section{Thermodynamic phase transition to quantum phase transition crossover}

\begin{figure}
\centering
   \includegraphics[scale=0.9]{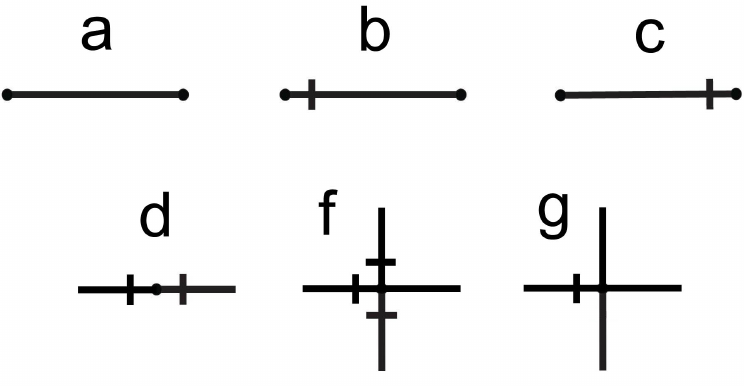}
   \caption{\textbf{The diagrammatic representation of the components of the Green functions matrix.}
   \textbf{a}: Keldysh Green function, $G^K$; \textbf{b}; advanced Green function, $G^A$;
   \textbf{c}: retarded Green function, $G^R$.}
   \label{GRF}
\end{figure}

Since quantum mechanics is an inherently dynamic theory, the dynamic Keldysh--Schwinger approach enables us to naturally describe the crossover from standard ``thermodynamic" phase transition to quantum phase transition at low temperatures.
Let us consider the system with the Lagrangian:
\begin{equation*}
    \mathcal{L}=\xi(\nabla \phi )^2+\sigma\phi^2+v\phi^4\,,
\end{equation*}
where $\phi $ is the scalar order parameter field, which obeys the Bose statistics.
Suppose that $v>0$ is a constant, and $\sigma $ is some parameter controlling the system state.
The system interacts with the heat bath having the temperature $T$. If $d>2$ then the system undergoes second order phase transition at $\sigma =0$.
The generating functional (partition functional) for the system is
\begin{gather*}
1=\mathcal{N}'\int \mathfrak{D}\phi^{cl} \mathfrak{D}\phi^{q}\exp \left[-\int\limits_{-\infty}^{\infty}\mathrm{d}\omega\int\limits_V\dfr{\mathrm{d}V}{V}\,\left\{\bar\phi\hat G'^{-1}\bar\phi+
2v\left(\phi^{cl}{\phi^{q}}^3+\phi^{q}{\phi^{cl}}^3\right)\right\}\right],
\end{gather*}
where $G'^{-1}$ is the massive $\phi$-field inverse Green function operator, which in reciprocal space is
\begin{equation}\label{GF}
\hat G'^{-1}=\tau^3\left[ \begin{array}{cc} 0 & \displaystyle\Box+\sigma +i \tau^{-1}\omega \\[12pt]
\displaystyle\Box+\sigma -i \tau^{-1}\omega & \displaystyle \displaystyle\tau^{-1}\omega \,\mbox{coth}\left({\hbar\omega}/{2k_{\mathrm{\scriptscriptstyle B}}T}\right) \end{array}\right].
\end{equation}
The diagrams representing the Green functions and non-linear vertices are shown in Fig.\,\ref{GRF}.

\begin{figure}
\centering
   \includegraphics[scale=0.7]{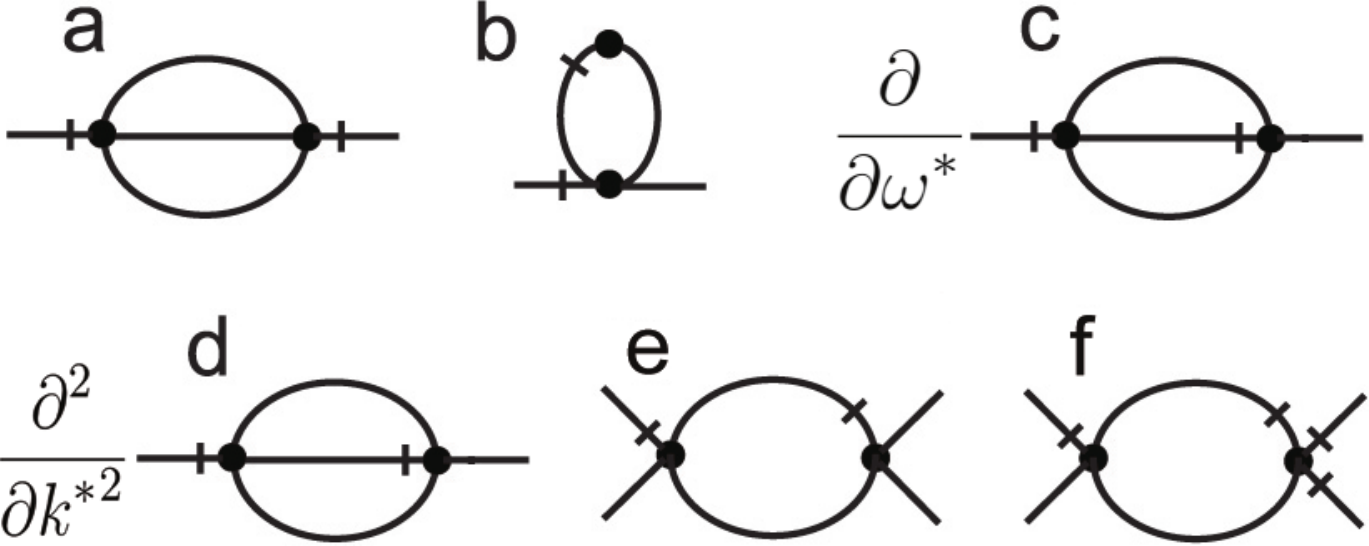}
   \caption{ \textbf{The diagrammatic representation of the contributions to the renormalization of the  vertices.}
   \textbf{a}: Renormalization of $G^{-1}_K$, \textbf{b}: Renormalization of $\sigma$, \textbf{c} and \textbf{d}: renormalization of $v$.}
   \label{f3}
\end{figure}

Near the critical point of a continuous phase transition, the critical dynamics this system can be described in terms of the renormalization group approach. The main logarithmically divergent contributions to the renormalized vertices are shown in Fig.\,\ref{f3}. To illustrate the divergences calculation, let us consider b, e and f terms. Each of these contributions contains the loop consisting only of one advanced (or retarded) Green function and one Keldysh Green function\,\cite{Vasin,VVR}. Thus, each of them is proportional to the integral
\begin{gather*}
I_b= \dfr{\hbar {V_{\bf k}}^{-1}}{2k_bT}\int\limits^{\lambda k^*}_{k^*} \mathrm{d}\omega\mathrm{d}{\bf k}\dfr{\tau^{-5}\omega\coth\left({\hbar\omega}/{2k_{\mathrm{\scriptscriptstyle B}}T}\right)}{(\Box+\sigma -i\tau^{-1}\omega)((\Box+\sigma)^2+ \tau^{-2}\omega^2)},
\end{gather*}
where $k^*$ is the 4-impulse approaching to $0$, and $\lambda $ is the regularization parameter.
When $\sigma \to 0$, i.e. in the long-wavelength limit corresponding to the critical dynamics near the QCP, the contribution of this loop is
\begin{gather}
\sim \dfr{\hbar}{2k_bT}\int\limits^{\lambda\omega^*}_{\omega^*} \mathrm{d}\omega^{1+d/z}\dfr{\omega\coth\left({\hbar\omega}/{2k_{\mathrm{\scriptscriptstyle B}}T}\right)}{\omega^3}.
\label{A0}
\end{gather}
Going to the dimensionless variable, one reduces expression\,(\ref{A0}) to
\begin{gather}
\sim \int\limits^{\lambda x^*}_{x^*} \mathrm{d}x^{1+d/z}\dfr{x\coth\left(x\right)}{x^3},
\end{gather}
where $x={\hbar\omega}/{2k_{\mathrm{\scriptscriptstyle B}}T}$.
One sees then that it is the factor $\coth(x)$ that determines the system's critical behavior, since this integral diverges logarithmically when $\coth(x)\sim x ^{1-d/z}$.

\begin{figure}
\centering
   \includegraphics[scale=1.2]{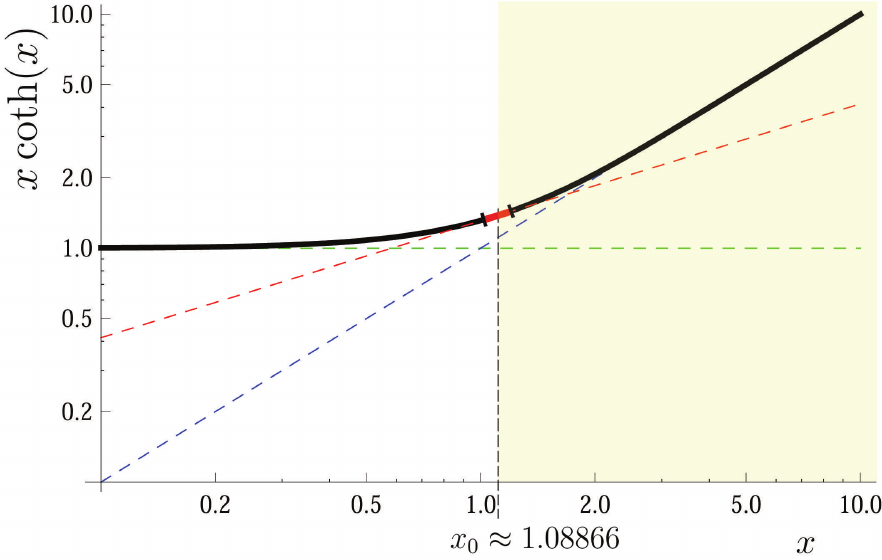}
   \caption{ The thick black line is the log--log plot of the $x\coth(x)$ function. The red dashed line is the tangent to $x\coth(x)$ function in $x=x_0$ point, the blue dashed line is the tangent to $x\coth(x)$ function at $x\gg 1$, and the green dashed line is the tangent to $x\coth(x)$ function at $x\to 0$ point. The straights in the log--log plot correspond to exponential functions $\propto x^{\Lambda}$ in linear coordinates. Thus, in neighborhood of some point $x=x^*$ the $x\coth(x)$ function can be approximated by the exponent function $\exp[\Lambda(x^*)\ln x+A(x^*)]$. In the point $x=x_0$ the $x\coth(x)$ function (the thick red linear segment) is approximated by the $\exp[\Lambda(x_0)\ln x+A(x_0)]$ with $\Lambda(x_0)\approx 1/2$. The yellow area corresponds to $x>x_0$ values, at which $\Lambda >1/2$. Here the fluctuation theory of phase transitions does not work in 3D system. }
   \label{LogPlot}
\end{figure}

In order to reduce the renormalization procedure to the standard form one should approximate the function $x\coth(x)$ by an exponential function, $ x^{\Lambda (x)}$.
Note that the critical dynamics considers a system in $\omega \to 0$ limit. However, this theoretical limit is practically unreachable. The natural limits are the observation time or the scale of zero-point fluctuations time in the quantum case.
Let us consider this approximation near to the lower limit of the frequency scale, $\omega = \omega_0$, that is relevant for critical dynamics. In the Fig.\,\ref{LogPlot} the $x\coth(x)$ function is shown in logarithmic coordinates. The sought exponential approximation of this function in some point $x=x^*$ is the tangent at this point. It is $\ln (x\coth(x))=\Lambda (x^*)\ln x +A(x^*)$, where $\Lambda (x^*)=\left.\partial \ln(x\coth(x))/\partial \ln x \right|_{x=x^*}$, and $A(x^*)=\ln\left({x^*}^{1-\Lambda (x^*)}\coth(x^*)\right)$.
Besides the renormalization procedure involves the integration just over narrow momentum interval. Therefore at $x\approx x^*$ the approximation with the good accuracy assumes the following form
\begin{gather}
\left. x\coth(x)\right|_{x\approx x^*}\propto x^{\Lambda (x^*)}\exp A(x^*),
\end{gather}
where
\begin{gather*}
\Lambda (x^*) = 1-2x^*\mbox{csch}(2x^*).
\end{gather*}
This is a single-valued exponential approximation of the considered function at the given point.
As a result, the loop contribution is proportional to
\begin{gather}
\int\limits^{\lambda\omega^*}_{\omega^*} \mathrm{d}\omega^{1+d/z}\omega^{\Lambda-3}.
\label{A2}
\end{gather}
From which it follows that the fluctuations are relevant as long as $2-d/z-\Lambda=(d^+_c-D_{eff})/z>0$. The effective dimensionality of the system is $D_{eff}=d+z\Lambda $. One can see that in 3D system at temperatures where $\Lambda >1/2$, the fluctuation corrections becomes non-relevant. Then the phase transition is well described by the mean field theory, and the critical exponents are equal to their mean-field theory quantities \cite{VVR}. The $\Lambda =1/2$ condition corresponds to $x=x_0\approx 1.08866$. At larger $x$ the critical exponents equal to the mean-field theory ones, and the system's dynamics becomes adiabatic.

Since the fluctuations frequency is limited from below by the zero point fluctuations frequency, $\omega_0$, one can estimate the temperature of the crossover between thermal and quantum critical dynamics regimes as $T_0={\hbar\omega_0}/{2k_{\mathrm{\scriptscriptstyle B}}x_0}$. This is the finite value in 3D case.
At the same time, in the 2D system the fluctuation corrections are relevant at all $x$, and the critical exponents approach to their mean--field values holds down to the technical limit of temperature measurements.

\section{Critical exponents calculation for 3D case}

In order to calculate the critical exponents in 3D, one can use the standard renormalization group protocol with Wilson $\varepsilon$-expansion.
According to this approach, in systems with $d<d^+_c$ the divergent contributions to vertices renormalization  are reduced to the logariphmically divergent ones by the formal replacement of the spatial dimensionality $d\to d+\varepsilon$:
\begin{gather*}
\propto\int \dfr{\mathrm{d}^{d+\varepsilon}k}{k^{d^+_c}}.
\end{gather*}
The $\varepsilon=d^+_c-d$ parameter is considered small, and the critical exponents are calculates as corrections to the exponent values of mean-field theory.

For considered system the upper-critical dimensionality is $d_c^+=4$, but the spatial dimensionality is $d=3$. In the classical limit $D_{eff}=d$. Therefore, calculations of the loops contribution are carried out for the system with dimensionality $d_c^+-\varepsilon$, where $\varepsilon=d_c^+-D_{eff}$. In spite of the fact that $\varepsilon =d_c^+-d=1$ it is belived small in order to these contributions become logarithmic. This is very strong assumption. However, in our case when the system approaches to the quantum limit, $\varepsilon$ really becomes small, since it can be represented as $\varepsilon (\Lambda) = d^+_c-D_{eff}(\Lambda)=1-2\Lambda$.
Then in the framework of the $\varepsilon$-expansion for considered $\varphi^4$--model the critical exponents are written as follows:
\begin{gather}
\beta\approx\dfr12-\dfr{1}{6}\varepsilon=\dfr12-\dfr{1}{6}\left(1-2\Lambda\right),\quad \nu \approx \dfr12+\dfr{1}{12}\varepsilon=\dfr12+\dfr{1}{12}(1-2\Lambda),\\
\alpha\approx\dfr1{6}\varepsilon=\dfr1{6}(1-2\Lambda),\quad \gamma\approx 1+\dfr1{6}\varepsilon=1+\dfr1{6}(1-2\Lambda).
\end{gather}
Note the closer $\Lambda $ to $1/2$ the more precisely these values become.
\begin{figure}
\centering
   \includegraphics[scale=1.2]{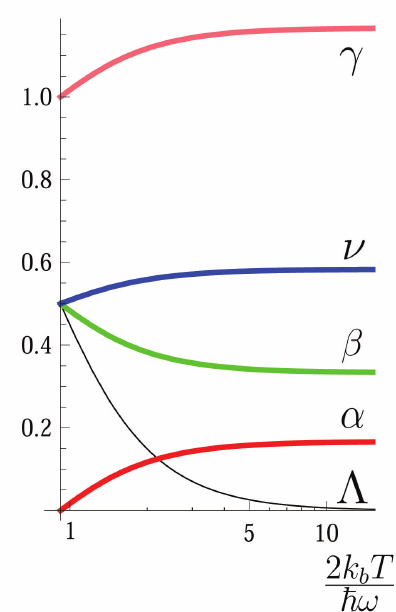}
   \caption{ Theoretical dependence of $\Lambda $ (black line), and the critical exponents: $\beta $ (green line), $\alpha$ (red line), $\nu $ (blue line), and $\gamma $ (pink line) on the $1/x={2k_{\mathrm{\scriptscriptstyle B}}T}/{\hbar\omega}$ ratio for 3D $\phi^4$-model. When $\Lambda $ becomes to be equal to $1/2$ and, accordingly, $D_{eff}=4$, then all exponents take the mean-field values.}
   \label{CE}
\end{figure}
As a result, near to the quantum limit, the critical exponents depend on the temperature through the $\Lambda(T)$ dependence (see Fig.\,\ref{CE}). This can look surprising but is observed in experiments.

\section{Crossover to the adiabatic regime}

Let us consider the renormalization of the dissipative parts of the Lagrangian close to the quantum critical point. The dissipative term in $[G^{-1}]^{A}$ and  $[G^{-1}]^{R}$ is $i\tau^{-1}\omega$.
The main contribution to the renormalization of this term corresponds to the {\bf c} graph in Fig.\,\ref{f3} and has the following form:
\begin{multline*}
I_c= \lim\limits_{\omega*\to 0}\dfr{\partial}{\partial\omega^*}\dfr{\hbar {V_{\bf k}}^{-2}}{2k_bT}\iint\limits^{\lambda k^*}_{k^*} \mathrm{d}\omega\mathrm{d}{\bf k}\mathrm{d}\omega'\mathrm{d}{\bf k}'\dfr{\tau^{-10}(\omega+\omega^*)\coth\left({\hbar(\omega+\omega^*)}/{2k_bT}\right)}{((\Box+\sigma)^2+ \tau^{-2}(\omega+\omega^*)^2)}\times\\
\dfr{(\omega+\omega')\coth\left({\hbar(\omega+\omega')}/{2k_bT}\right)}{(\Box'+\sigma+ i\tau^{-1}\omega')((\Box''+\sigma)^2+ \tau^{-2}(\omega'+\omega)^2)}.
\end{multline*}
Taking into account the above approximation $x\coth x\to x^{\Lambda}$ near the cutoff frequency and performing differentiation over $\omega $ one can find that this expression is proportional to
\begin{gather*}
\int\mathrm{d}\omega\,\omega^{d-5+2\Lambda}.
\end{gather*}
This integral diverges while $d-4+2\Lambda\leqslant 0$. In case of a 3D system this condition is $\Lambda\leqslant 1/2$. Otherwise it becomes non-relevant and $i\tau^{-1}\omega$ term vanishes from the Lagrangian. By analogy one can show that ${\bf k}^2$-term also vanishes.

In formal generalization the Keldysh term in (\ref{GF}) which is proportional to $\omega^{\Lambda}$ are renormalized by the term which divergence is caused by the properties of following integral:
\begin{multline*}
\partial^{\Lambda}_{\omega}I_a= \lim\limits_{\omega*\to 0}\dfr{\partial^{\Lambda}}{\partial{\omega^*}^{\Lambda}}\dfr{\hbar {V_{\bf k}}^{-2}}{2k_bT}\iint\limits^{\lambda k^*}_{k^*} \mathrm{d}\omega\mathrm{d}{\bf k}\mathrm{d}\omega'\mathrm{d}{\bf k}'\dfr{\tau^{-10}(\omega+\omega^*)\coth\left({\hbar(\omega+\omega^*)}/{2k_bT}\right)}{((\Box+\sigma)^2+ \tau^{-2}(\omega+\omega^*)^2)}\times\\
\dfr{\omega'\coth\left({\hbar\omega'}/{2k_bT}\right)
(\omega+\omega')\coth\left({\hbar(\omega+\omega')}/{2k_bT}\right)}{((\Box'+\sigma)^2+ \tau^{-2}\omega'^2)((\Box''+\sigma)^2+ \tau^{-2}(\omega'+\omega)^2)},
\end{multline*}
where $\partial^{\Lambda}_{\omega}$ means the fractional derivation over $\omega$. One can see that
\begin{gather*}
\partial^{\Lambda}_{\omega^*}I_a\propto \int\mathrm{d}\omega\,\partial^{\Lambda}_{\omega}\omega^{d-5+3\Lambda}\propto
\int\mathrm{d}\omega\,\omega^{d-5+2\Lambda},
\end{gather*}
and for a 3D system this correction becomes non-relevant when $\Lambda>1/2$.

Thus, all fluctuation corrections to the Lagrangian becomes non-relevant at $\Lambda>1/2$, and the theory transforms to the mean field one. In this case the system critical dynamics becomes adiabatic, and critical behaviour of the system corresponds to quantum $d+1$ dimensionality. This agrees with the above critical exponents calculation.

\section{Conclusions}

To summarize, at temperatures approaching to zero, the critical exponents of the system converge to those of the mean-field theory.  This crossover occurs continuously with the temperature decrease and contrasts the behaviour which one would have expected from the orthodox viewpoint, in which the universality class that determines the value of the exponent does not vary with the temperature. However, as we have shown, the critical exponents change only when close enough to the quantum critical point. The reason for that is that the critical exponents depend on the nature of the critical fluctuations. Strong thermal fluctuations have the ``white'' noise spectrum whereas quantum fluctuations are endowed with the ``color'' noise.  Thus the temperature acts here as a parameter which defines the fluctuation type and their spectrum, hence influencing on the critical exponents.  At high temperatures, $T\gg {\hbar\omega_0}/{2k_{\mathrm{\scriptscriptstyle B}}x_0}$, the theory effective dimensionality is equal to the spatial dimensionality $D_{eff}=d=3$, therefore, the exponents are the ones characteristic to the three-dimensional classical system. At low temperatures the effective system dimensionality in renormalization procedure reaches the upper critical value, $D_{eff}=d+\Lambda z\geqslant d_c^*$. As a result, the critical exponents become equal to the mean-field theory ones. Note that the system's universality class, defined by the system symmetry, remaining intact.
%Thus, the continuous change in the critical exponents reflects the increase in the effective dimensionality of the system near critical point caused by the crossover from the thermal fluctuations mode, which is ``white'' noise, to the quantum fluctuations regime, which is ``blue'' noise, i.e. the key factor here is the fluctuation spectrum change.

\section*{Acknowledgments}
The work at Argonne (V.M.V) was supported by the U.S. Department of Energy,
Office of Science, Basic Energy Sciences, Materials Sciences
and Engineering Division.

\section{Appendices}

\subsection*{Appendix I}

An extremely useful tool for us is the algebra of bosonic coherent
states. In this formulation the state of a many body system can be presented with the bosonic annihilation and creation operators, $\hat b$ and $\hat b^{\dag}$, which operate in the space of the boson occupation numbers $n$ in the following way:
\begin{gather*}
\hat b|n\rangle =\sqrt{n}|n-1\rangle, \quad \hat b^{\dag}|n\rangle =\sqrt{n+1}|n+1\rangle
\end{gather*}
where the number states $|n\rangle$ form a complete orthonormal basis: $\langle n|n'\rangle=\delta_{n,n'}$, and $\sum\limits_n |n \rangle \langle n|=\hat 1$. By acting on an arbitrary basis state,
one may check the following relations
\begin{gather*}
\hat b^{\dag}\hat b|n\rangle =n|n\rangle, \quad \hat b\hat b^{\dag}|n\rangle =\sqrt{n+1}|n\rangle,\quad [\hat b,\,\hat b^{\dag}]=\hat 1
\end{gather*}
Therefore, the coherent state of the system, parameterized by a complex number $\phi $, is defined as eigenstates of the annihilation operator with the eigenvalue $\phi $
\begin{gather*}
\hat b|\phi\rangle =\phi|\phi\rangle, \quad \langle \phi | \hat b^{\dag} =\phi^*\langle \phi |
\end{gather*}
where the star denotes complex conjugation. Then, the matrix elements in the coherent state basis of any normally ordered operator $\hat H(\hat b^{\dag},\,\hat b)$ are given by
\begin{gather*}
\langle\phi|\hat H(\hat b^{\dag},\,\hat b)|\phi'\rangle =\hat H(\phi^*,\,\phi')\langle\phi|\phi'\rangle
\end{gather*}
One can check that the following linear superposition of the pure number states:
\begin{gather*}
|\phi\rangle=\sum\limits_{n=0}^{\infty}\dfr{\phi^n}{\sqrt{n!}}|n\rangle=\sum\limits_{n=0}^{\infty}\dfr{\phi^n}{\sqrt{n!}}(\hat b^{\dag})^n|0\rangle=e^{\phi \hat b^{\dag}}|0\rangle
\end{gather*}
where $|0\rangle $ is the vacuum state, is the required eigenstate of the operator $\hat b$. Upon Hermitian conjugation, one
finds $\langle \phi|=\sum\limits_{n}\langle n|{\phi^*}^n/\sqrt{n!}$.

Note that the coherent states are not mutually orthogonal: their set forms an overcomplete basis. The overlap of two coherent states is given by
\begin{gather*}
\langle \phi|\phi'\rangle=\sum\limits_{n,n'=0}^{\infty}\dfr{{\phi^*}^n{\phi'}^{n'}}{\sqrt{n!n'!}}\langle n|n'\rangle=\sum\limits_{n=0}^{\infty}\dfr{(\phi^*\phi')^n}{\sqrt{n!}}=e^{\phi^*\phi'}
\end{gather*}
where we employed the orthonormality of the pure number states. One may express resolution of unity in the coherent states basis. It takes the following
form:
\begin{gather*}
\hat 1=\int\mathfrak{D}\phi e^{-|\phi|^2} |\phi\rangle\langle \phi |
\end{gather*}
where $\mathfrak{D}\phi=\mathfrak{D}(\Re \phi)\mathfrak{D}(\Im \phi)$ denotes the functional integration over real and imaginary parts of $\phi$ field.
The action of the time derivation operator on the state function, important for us, is
\begin{multline*}
\partial_t|\phi\rangle=\sum\limits_{n=0}^{\infty}\dfr{\partial_t\phi^n}{\sqrt{n!}}|n\rangle=
(\partial_t\phi)\sum\limits_{n=0}^{\infty}\dfr{n\phi^{n-1}}{\sqrt{n!}}|n\rangle= \\
(\partial_t\phi)\sum\limits_{n=0}^{\infty}\dfr{\sqrt{n}\phi^{n-1}}{\sqrt{(n-1)!}}|n\rangle=
(\partial_t\phi)\hat b^{\dag}\sum\limits_{n=0}^{\infty}\dfr{\phi^{n-1}}{\sqrt{(n-1)!}}|n-1\rangle=\\
(\partial_t\phi)\hat b^{\dag}|\phi\rangle= \phi^*(\partial_t \phi )|\phi\rangle
\end{multline*}
Therefore,
\begin{gather*}
\langle \phi|\partial_t|\phi\rangle=\phi^*(\partial_t \phi )\langle \phi|\phi\rangle =\phi^*(\partial_t \phi )e^{|\phi|^2}
\end{gather*}

\subsection*{Appendix II}

Upon dividing the time interval $(t',\,t'')$ into the infinitesimal parts $\delta t$, the probability for the transition $\phi_{t'}\to\phi_{t''}$ becomes
\begin{gather*}
\langle \phi_{t''}|\mathcal{\hat U}|\phi_{t'}\rangle= \int\mathfrak{D}\phi_{t''-\delta t}\dots\int\mathfrak{D}\phi_{t'+\delta t}\langle \phi_{t''}|\mathcal{\hat U}_{\delta t}| \phi_{t''-\delta t}
\rangle \times\\
\langle \phi_{t''-\delta t}|\mathcal{\hat U}_{\delta t}| \phi_{t''-2\delta t} \rangle \dots \langle \phi_{t'+2\delta t}|\mathcal{\hat U}_{\delta t}| \phi_{t'+\delta t} \rangle \langle \phi_{t'+\delta
t}|\mathcal{\hat U}_{\delta t}| \phi_{t'} \rangle ,
\end{gather*}
where $\mathcal{\hat U}_{\delta t}$ is the evolution operator on the time interval $\delta t$. The $| \phi_{j-1}\rangle$, and $| \phi_{j}\rangle$ states are not coherent. Therefore, the evolution
operator elements are given by
\begin{gather*}
\langle \phi_j|\mathcal{\hat U}_{\delta t}| \phi_{j-1}\rangle \equiv
\langle \phi_j|e^{-i\hat H\hbar^{-1}\delta t}| \phi_{j-1}\rangle \approx \\
\langle \phi_j|\hat 1-{i\hbar^{-1}\hat H\delta t}| \phi_{j-1}\rangle =
\langle \phi_j| \phi_{j-1}\rangle (1-{i\hbar^{-1}H_j\delta t}) =\\
\langle \phi_j| \hat 1-\delta t\partial_t|\phi_{j}\rangle (1-{i\hbar^{-1}H_j\delta t})=\\
\langle \phi_j|\phi_{j}\rangle e^{-\delta t\phi_j^*\partial_t\phi_{j}}e^{-i\hbar^{-1}H_j\delta t}=e^{-\phi_j^*\partial_t\phi_{j}\delta t-i\hbar^{-1}H_j\delta t}
\end{gather*}
(see Appendix I), and in the continuous limit we have
\begin{gather*}
\langle \phi_{t''}|\mathcal{\hat U}|\phi_{t'}\rangle=
 \int \mathfrak{D}\phi\exp \left[i\int\limits_{t'}^{t''}\mathrm{d}t\,\left(\hbar^{-1}\mathcal{L}(\phi)+i\int\limits_V\dfr{\mathrm{d}V}V
\,\phi\partial_t\phi\right) \right].
\end{gather*}
This expression can be rewritten as
\begin{gather*}
\langle \phi_{t''}| \mathcal{\hat U}|\phi_{t'}\rangle=
\int \mathfrak{D}\phi \exp \left[i\int\limits_{t'}^{t''}\mathrm{d}t\int\limits_V\dfr{\mathrm{d}V}V\,\phi G^{-1}\phi \right],
\end{gather*}
where\,\cite{KAMENEV}

\begin{gather*}
G^{-1} = \hbar^{-1}\mu\partial_t^2-\hbar^{-1}\xi \nabla^2 + i\partial_t\,.
\end{gather*}

\subsection*{Appendix III}

We use the abbreviated notation, which means operations with complex variables:
\begin{multline*}
\int \mathfrak{D}z \exp\left[-zAz +az\right]=\\
\iint \mathfrak{D}(\Re z)\mathfrak{D}(\Im z)\exp\left[-\iint \mathrm{d}{\bf x}\mathrm{d}{\bf y}\,z_{\bf x}^*A_{\bf x-y}z_{\bf y}+ \right. \\ \left.
\int \mathrm{d}{\bf x} \,a_{\bf x}^*z_{\bf x}+\int\mathrm{d}{\bf x}\,z_{\bf x}^*a_{\bf x}\right]\\=\dfr{1}{\det A}\exp\left[\iint\mathrm{d}{\bf x}\mathrm{d}{\bf y}\, a_{\bf x}^*A^{-1}_{\bf x-y}a_{\bf
y}\right]=\dfr{\exp\left[a^*A^{-1}a\right]}{\det A}
\end{multline*}
where $\int \mathfrak{D}z$ denotes the functional integration over $z$ field, and $z^*$ field is the complex conjugate to $z$.

\end{document}